\documentclass[sigconf]{acmart}
\usepackage{enumitem}
\usepackage{multirow}
\usepackage{graphicx}
\usepackage{subcaption}
\usepackage{arydshln}
\usepackage{tikz} 
\usepackage{tabularx}
\usepackage{booktabs}
\usepackage{float}
\usepackage[absolute,overlay]{textpos}
\newcommand{\ourmodel}{S-GRec}
\AtBeginDocument{%
  }

\setcopyright{none}
\renewcommand\footnotetextcopyrightpermission[1]{}
\begin{document}

\begin{textblock*}{\paperwidth}(1.5cm, 1.2cm)
\includegraphics[height=1.4cm]{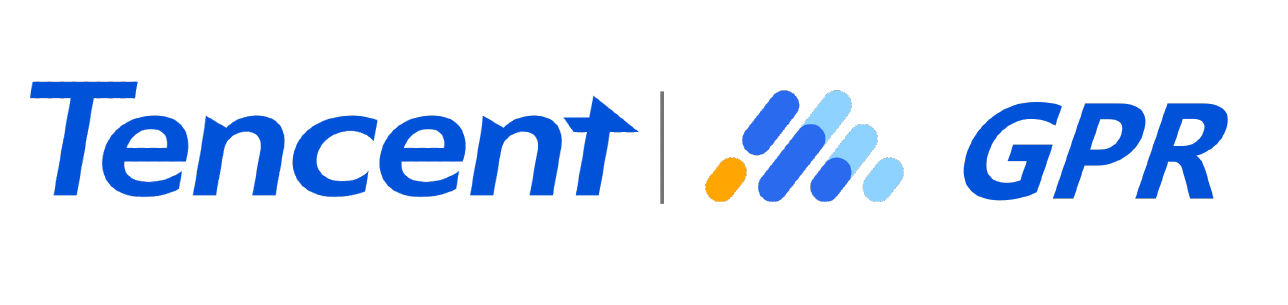}
\end{textblock*}

\setlength{\abovedisplayskip}{2pt}
\setlength{\belowdisplayskip}{2pt}
\setlength{\abovedisplayshortskip}{2pt}
\setlength{\belowdisplayshortskip}{2pt}

\setlength{\textfloatsep}{3pt}   
\setlength{\intextsep}{3pt}      
\setlength{\abovecaptionskip}{2pt}
\setlength{\belowcaptionskip}{0pt}

\title{S-GRec: Personalized Semantic-Aware Generative Recommendation with Asymmetric Advantage}

\author{Jie Jiang$^{*}$, Hongbo Tang$^{*}$, Wenjie Wu$^{*}$, Yangru Huang$^{*}$, Zhenmao Li$^{*}$, Qian Li, Changping Wang, \\Jun Zhang$^{\dagger}$, Huan Yu}
\affiliation{
  \institution{Tencent Inc., China}
  \country{
    \{zeus, hardytang, yonewu, yarayrhuang, zoomli, kathieqli, terracewang, neoxzhang, huanyu\}@tencent.com }
}
\thanks{$^{*}$Equal contribution. \\ $^{\dagger}$Corresponding author.}

\renewcommand{\shortauthors}{Jiang et al.}

\begin{abstract}

Generative recommendation models sequence generation to produce items end-to-end, but training from behavioral logs often provides weak supervision on underlying user intent. Although Large Language Models (LLMs) offer rich semantic priors that could supply such supervision, direct adoption in industrial recommendation is hindered by two obstacles: semantic signals can conflict with platform business objectives, and LLM inference is prohibitively expensive at scale. This paper presents \textbf{S-GRec}, a semantic-aware framework that decouples an online lightweight generator from an offline LLM-based semantic judge for train-time supervision. S-GRec introduces a two-stage \textbf{Personalized Semantic Judge (PSJ)} that produces interpretable aspect evidence and learns user-conditional aggregation from pairwise feedback, yielding stable semantic rewards. To prevent semantic supervision from deviating from business goals, \textbf{Asymmetric Advantage Policy Optimization (A2PO)} anchors optimization on business rewards (e.g., eCPM) and injects semantic advantages only when they are consistent. Extensive experiments on public benchmarks and a large-scale production system validate both effectiveness and scalability, including statistically significant gains in CTR and a \textbf{1.19\%} lift in GMV in online A/B tests, without requiring real-time LLM inference.

\end{abstract}

\keywords{Generative Recommendation, Large Language Models, Policy Optimization, Metric Alignment, Semantic-aware Modeling}

\maketitle

\section{Introduction}

Recommendation systems have evolved from the conventional retrieval--ranking paradigm into generative recommendation \cite{hou2025survey,xiao2025survey}, where recommendation is formulated as sequence generation to enable end-to-end construction of item lists and richer modeling of temporal and contextual dependencies in user behavior. This paradigm is increasingly adopted across industrial applications such as e-commerce \cite{hao2025oxygenrec,tang2025interactive,chen2025onesearch,guo-OneSug-2025}, short-video platforms \cite{OneRec-2025, OneRec-v2-arxiv-2025}, social media feeds \cite{HSTU-ICML-2024}, POI recommendation \cite{wei2025oneloc} and online advertising \cite{zhang2025gpr,zheng-EGAV2-2025, qiu2025one}. Its practical significance is reflected in the fact that recommendation outputs simultaneously shape user experience and directly drive key business metrics, including Click-Through Rate (CTR), Conversion Rate (CVR), and Gross Merchandise Volume (GMV). However, in real-world deployments, user preferences are often implicit and fine-grained, while behavioral feedback is noisy and biased. As a result, relying on behavior sequences for generative learning tends to induce superficial correlation fitting, making it difficult to infer why users prefer certain items. More critically, it struggles to maintain stable alignment with the platform's business objectives.

\begin{figure}[t]
\centering
\begin{subfigure}[t]{0.27\columnwidth}
\centering
\includegraphics[height=4cm, keepaspectratio]{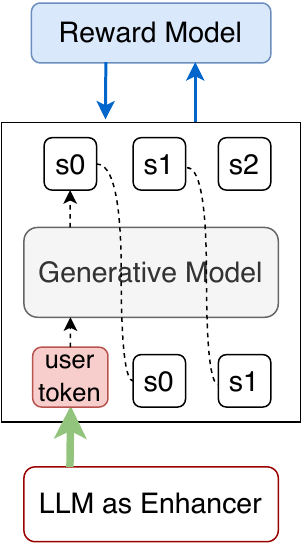}
\caption{}
\label{fig:introduction_a}
\end{subfigure}
\hfill
\tikz\draw [gray, densely dashed] (0,0) -- (0,4); 
\hfill
\begin{subfigure}[t]{0.32\columnwidth}
\centering
\includegraphics[height=4cm, keepaspectratio]{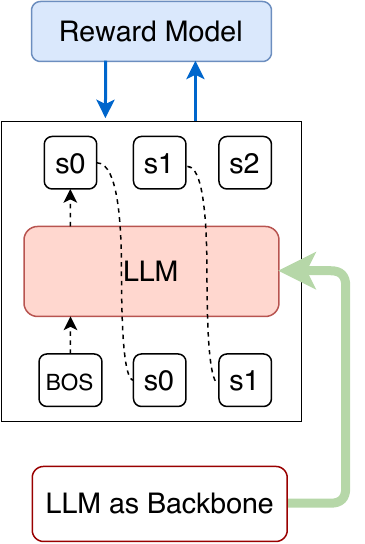}
\caption{}
\label{fig:introduction_b}
\end{subfigure}
\hfill
\tikz\draw [gray, densely dashed] (0,0) -- (0,4); 
\hfill
\begin{subfigure}[t]{0.34\columnwidth}
\centering
\includegraphics[height=4cm, keepaspectratio]{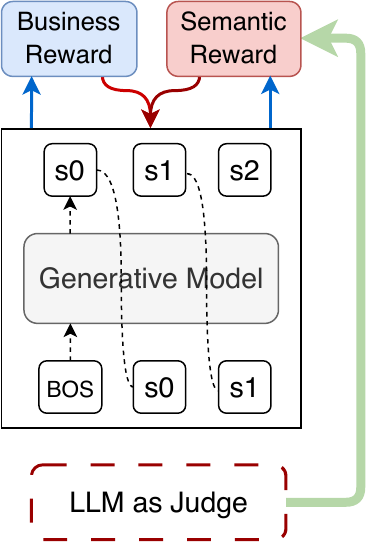}
\caption{Ours}
\label{fig:introduction_c}
\end{subfigure}
\caption{LLM applications in different ways. The dashed red line indicates LLM use only during training for our method, while the other two approaches require it in both training and inference.
}
\label{fig:introduction}
\end{figure}

With the rapid progress of Large Language Models (LLMs) \cite{achiam2023gpt, DeepSeekR1-Nature-2025, qwen2.5}, their powerful semantic understanding and reasoning capabilities have sparked significant interest in LLM-integrated recommendation systems \cite{REC-R1-2025, yi2025recgpt}. Current research primarily follows two paradigms:
(i) LLMs as semantic enhancers: as shown in Figure~\ref{fig:introduction_a}, these methods leverage LLMs to deeply encode textual features (e.g., user profiles and item metadata) into high-quality semantic representations or discrete IDs. These enhanced features are then integrated into downstream recommenders to better capture content semantics and user intentions \cite{chen-HLLM-2024, yi2025recgpt-v2}. (ii) LLMs as primary generators: as shown in Figure~\ref{fig:introduction_b}, these approaches treat recommendation as a generative task, fine-tuning LLMs via instruction tuning or Parameter-Efficient Fine-Tuning (e.g., LoRA) to directly generate recommendation lists or predict item ratings \cite{TALLRec-RecSys-2023, GenRec-RecSys-2023}. While these approaches validate the potential of LLMs for recommendation, they face critical bottlenecks in the industrial deployment of generative recommendation.

The remaining deployment gap is primarily driven by two structural conflicts between LLMs and generative recommendation in industrial scenarios. The first is the \textit{Cost-Scalability Conflict}: While LLMs offer immense semantic power, they are burdened by high inference costs and substantial memory footprints. In contrast, production-scale recommenders necessitate millisecond-level latency and high-throughput scalability to handle millions of requests, creating a fundamental tension between model complexity and operational efficiency.
The second is the \textit{Objective Alignment Conflict}: Recommendation engines are engineered to optimize scenario-specific business metrics by modeling implicit user behaviors. Conversely, general-purpose LLMs are pre-trained for open-domain semantic generation \cite{LC-Rec-ICDE-2024, luo2025qarm}. Without a robust mechanism to translate ambiguous implicit preferences into explicit, business-aligned supervisory signals, these models struggle to maintain stable performance on platform-specific objectives. This raises a core question: \textit{How can we distill the semantic priors of LLMs into controllable preference supervision to train a generative recommender, while strictly maintaining alignment with business objectives?}

In this paper, we bridge these gaps by proposing \textbf{Semantic-Aware Generative Recommendation (S-GRec)}, a framework that reconciles LLM-driven semantic depth with the stringent requirements of industrial serving (see Figure~\ref{fig:introduction_c}). Our core philosophy is to decouple high-cost semantic reasoning from real-time inference, distilling it into a lightweight, business-aligned generator during training.
Specifically, S-GRec first introduces an LLM-based \textbf{Personalized Semantic Judge (PSJ)} to transform complex user--item semantic interactions into scalar preference signals. This is achieved via a two-stage pipeline: (i) \textit{Aspect-level Semantic Scoring}, which yields interpretable, fine-grained evidence to characterize candidate items; and (ii) \textit{User-conditional Preference Aggregation}, which employs GRPO-style optimization to learn context-aware importance weights from pairwise feedback. 
To prevent the distilled semantic supervision from conflicting with platform objectives, S-GRec further incorporates \textbf{Asymmetric Advantage Policy Optimization (A2PO)}. This mechanism fuses business rewards with PSJ-derived semantic rewards at the advantage level using consistency-aware asymmetric weighting. In addition, S-GRec adopts \textbf{sparse sampling} and queries PSJ on only a small subset of training instances. In summary, S-GRec delivers a robust framework that harmonizes high-quality semantic intelligence with platform-specific objectives, offering a scalable path to deploy generative recommenders without the prohibitive overhead of real-time LLM reasoning.

The main contributions of this paper are summarized as follows:
\begin{enumerate}[label=(\arabic*), leftmargin=*, itemsep=0pt, parsep=2pt]
    \item \textbf{LLM-as-Judge for Industrial Generative Recommendation.}
    We propose a decoupled framework that uses an LLM as an \emph{offline} semantic judge to provide controllable preference supervision for training, while keeping \emph{online} serving lightweight and free of LLM dependency.

    \item \textbf{Personalized Semantic Judge (PSJ).}
    We develop a two-stage judge adaptation pipeline that produces interpretable, aspect-level evidence and learns user-conditional aggregation from pairwise feedback, yielding stable scalar semantic rewards aligned with user preferences.

    \item \textbf{Asymmetric Advantage Policy Optimization (A2PO).}
    We propose an asymmetric,  advantage-level fusion objective that integrates PSJ semantic rewards with business rewards under sparse LLM querying, preventing conflicts with platform objectives at industrial cost.

    \item \textbf{Strong Offline and Online Results.}
    Extensive experiments on public benchmarks and large-scale online A/B tests demonstrate consistent improvements and practical impact, including a \textbf{1.19\%} GMV lift and a \textbf{1.55\%} GMV-Normal lift.
\end{enumerate}

\vspace{-1em}
\section{Related Works}

\subsection{Generative Recommendation}

Generative Recommendation (GR) formulates recommendation as conditional sequence generation, where an autoregressive model generates the target item from user context, collapsing retrieval and ranking into a single decoder \cite{TIGER-NeurIPS-2023,zhou2025onerec,OneRec-v2-arxiv-2025}. Prior work advances GR along two main directions: (i) \emph{scalable architectures and serving pipelines}, evolving from encoder-decoder to decoder-only and industrial-grade systems such as HSTU and the OneRec family \cite{HSTU-ICML-2024,zhou2025onerec,OneRec-v2-arxiv-2025}; and (ii) \emph{representation and tokenization schemes} that make item generation tractable at scale, often via quantization-based codes \cite{lee2022autoregressive,luo2025qarm,fsq}. GR has also been extended to richer signals and settings, including collaborative features, geographic context, and multi-behavior generation \cite{EAGER--KDD-2024,LETTER-CIKM-2024,wei2025oneloc,liu2024multi,xiao2025unger}. 
Despite these advances, a key bottleneck in industrial scenarios remains: GR training is still dominated by noisy behavioral logs and likelihood-oriented objectives, making it hard to inject semantic preference supervision while preserving business alignment \cite{kong2025sdpo,REC-R1-2025}. Recent methods introduce LLMs as generators, rerankers, or feature augmentation modules, but often incur online inference overhead or require heavy fine-tuning, limiting scalability and deployment. In contrast, this work adopts an \emph{LLM-as-Judge} paradigm, using the LLM as an \emph{offline} source of controllable semantic supervision. This enables scalable semantic guidance for GR training and motivates our domain-adapted two-stage judge and consistency-aware advantage-level fusion that maintains business alignment without adding serving-time cost.

\subsection{LLM-as-Judge and Preference Supervision}
LLM-as-Judge uses a strong language model to provide preference judgments or rubric-based scores, offering a scalable alternative to human labeling for evaluation and alignment. This paradigm is closely related to RLHF and RLAIF, where preference feedback, including AI-generated feedback, is used to optimize a target model \cite{bai2022constitutional,lee2023rlaif}. Prior work has studied the reliability of LLM judges and their agreement with humans \cite{zheng2023judging}, and developed specialized judge models or prompting frameworks to improve controllability and interpretability, including multi-aspect scoring and open-source judges \cite{liu2023g,kim2023prometheus,zhu2023judgelm}. Recent studies further investigate judge biases and calibration, such as position and verbosity bias, and propose mitigation or self-improvement strategies \cite{ye2024justice,yuan2024self,lai2025beyond}. LLM judges have also been used to scale preference datasets and distill supervision, for example in synthetic preference pipelines \cite{dubois2023alpacafarm,cui2023ultrafeedback}. In recommendation, LLM judges have mainly been used as external evaluators or offline modules. RecGPT and RecGPT-V2 employ LLM-based judging to assess recommendation quality and decompose evaluation into multiple aspects \cite{yi2025recgpt,yi2025recgpt-v2}. However, existing uses typically do not address a key industrial training issue: semantic judgments can be noisy and may conflict with business rewards, making naive reward mixing unstable. In contrast, this work treats the LLM as a domain-adapted semantic judge that produces structured preference signals and integrates them into generative recommendation training via advantage-level, consistency-aware fusion, aiming to improve semantic preference modeling while preserving business alignment under industrial cost constraints.

\vspace{-0.5em}
\section{Methodology}

\begin{figure*}[htbp]
    \centering
    \includegraphics[width=1\textwidth]{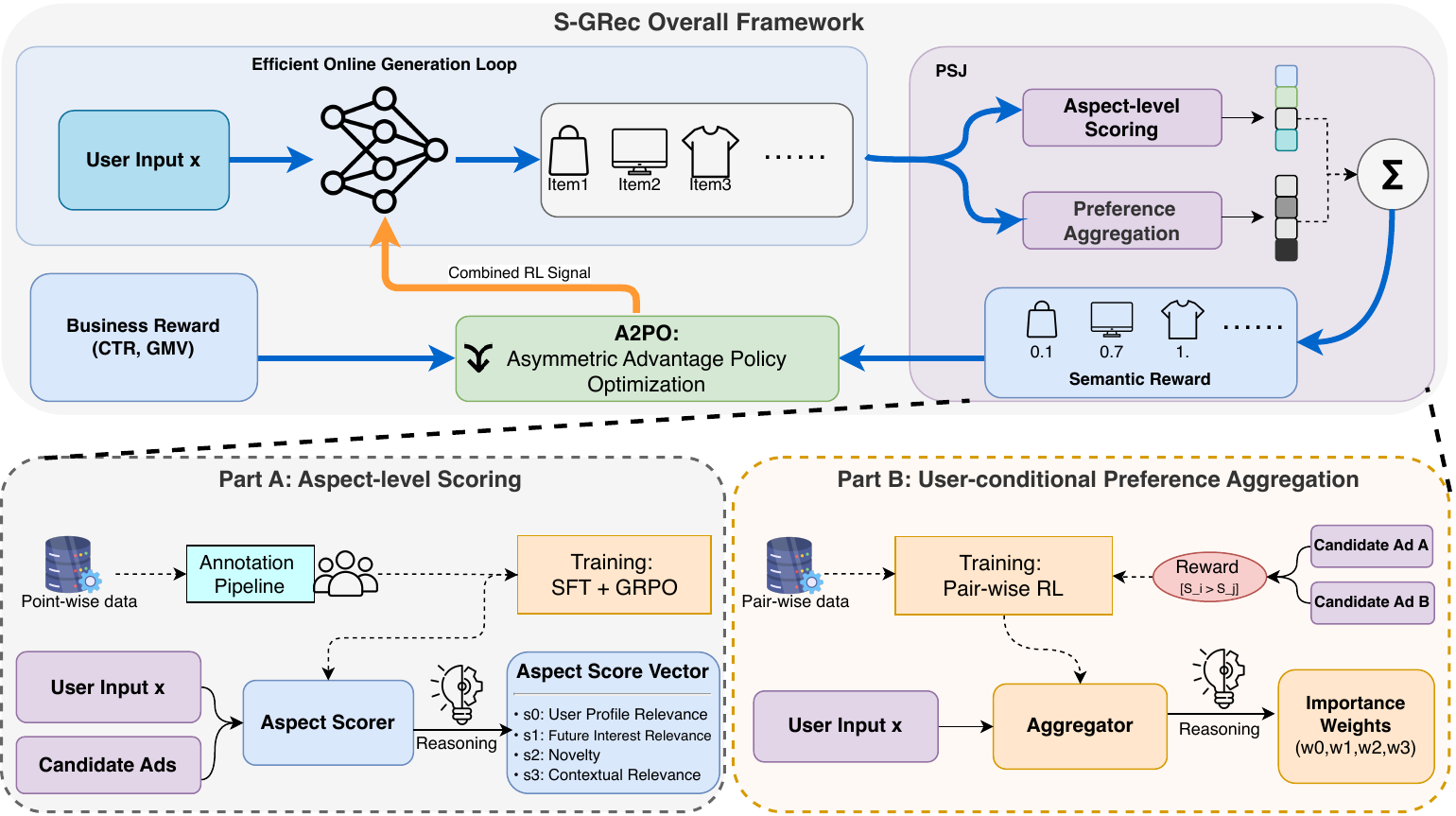}
        \caption{Overview of S-GRec. The offline PSJ produces semantic rewards via two-stage scoring (aspect-level evidence $\rightarrow$ user-conditional aggregation). A2PO fuses semantic and business rewards in the advantage space with consistency gating, training a lightweight generator without serving-time LLM inference.}
    \label{fig:overview}
    \vspace{-0.5em}
\end{figure*}

In this section, we first formalize the problem and then describe the two core components of S-GRec. First, we detail the Personalized Semantic Judge (PSJ) and its two-stage training pipeline for fine-grained content scoring and preference alignment, including the scoring rubrics and LLM-based supervision. We then introduce Asymmetric Advantage Policy Optimization (A2PO), which fuses LLM-derived semantic rewards with business signals in a unified objective. Figure~\ref{fig:overview} illustrates the overall architecture.

\subsection{Preliminary}
\label{sec:preliminary}

\paragraph{Generative recommendation with Semantic IDs \cite{TIGER-NeurIPS-2023}.}
We formulate recommendation as conditional sequence generation.
Given a unified input $x$ that encapsulates the user profile, behavioral history, and spatio-temporal context (e.g., time and location), each item (or ad) is represented by a fixed-length Semantic ID (SID) token sequence
$y=(y_1,\ldots,y_T)\in\mathcal{V}^{T}$.
An autoregressive policy $\pi_{\theta}$ models
\begin{equation}
\pi_{\theta}(y \mid x) = \prod_{t=1}^{T} \pi_{\theta}\!\left(y_t \mid x, y_{<t}\right),
\label{eq:gen_factorization}
\end{equation}
and a deterministic mapping $\mathrm{Map}(\cdot)$ maps a valid SID sequence $y$ to its corresponding item. Candidate items are obtained by generating multiple valid SIDs under the same $x$.

\paragraph{Group-based sampling for policy optimization \cite{DeepSeekR1-Nature-2025}.}
For each $x$, we sample a group of $G$ candidate sequences
$\mathcal{Y}(x)=\{y^{1},y^{2},\ldots,y^{G}\}$
from a rollout policy and evaluate each candidate with a scalar reward $r(x,y)$.
Following group-based policy optimization, we use \emph{group-relative} (standardized) advantages computed within $\mathcal{Y}(x)$ to reduce reward scale sensitivity and stabilize updates.

\paragraph{Why not off-the-shelf LLM-as-Judge.}

Direct application of general-purpose LLMs as judges is impractical for large-scale ad recommendation. Large models face prohibitive inference overhead, while smaller variants are often fragile, showing sensitivity to prompt phrasing and candidate list presentation. Moreover, off-the-shelf LLMs struggle with ad-domain alignment, failing to capture interest evolution or business constraints while suffering from judgment drift under finite context windows. This motivates the need for a structured, domain-stabilized judge.

\subsection{Personalized Semantic Judge (PSJ): Turning an LLM into a Reward Provider}
\label{sec:hsj}

\subsubsection{Problem Setup}

Let $x$ denote the comprehensive user input, including the profile, recent behavior sequence, and request-time signals (e.g., time and location when available).
Given a candidate ad $a=\mathrm{Map}(y)$, PSJ serves as a semantic reward provider for downstream recommendation optimization. 
For each $(x,a)$, PSJ returns:
(i) an interpretable aspect score vector $\mathbf{s}(x,a)\in\mathbb{R}^4$ over four semantic dimensions; and
(ii) a holistic semantic score $s_{\mathrm{hol}}(x,a)\in\mathbb{R}$ as a scalar semantic reward for ranking and RL.
The design follows a progressive principle: PSJ first learns decomposed aspect judgments, and then learns a user-conditional aggregation rule that maps the aspect evidence to a stable scalar preference.

\subsubsection{Part A: Aspect-level Semantic Scoring}
\vspace{4pt}

\paragraph{Point-wise fine-grained scoring.}
This task trains PSJ to output independent, interpretable judgments by decomposing each candidate ad into structured aspect scores.
Given a user input-ad pair $(x,a)$, PSJ predicts a discrete score vector $\mathbf{s}(x,a)$ over four semantic dimensions:
\begin{itemize}[leftmargin=*, itemsep=2pt, parsep=2pt]
    \item \textbf{User Profile Relevance}: alignment with the user's long-term and coarse-grained interests.
    \item \textbf{Future Interest Relevance}: alignment with short-term, fine-grained, and evolving interests.
    \item \textbf{Novelty}: relevance to derivative interests not yet covered by the user's exposure history.
    \item \textbf{Contextual Relevance}: suitability under request-time context such as time and location.
\end{itemize}
We adopt discrete labels to reduce annotation ambiguity and stabilize learning. The full scoring rubrics are provided in Table \ref{tab:fine_grained_scoring_specs}.

\paragraph{Point-wise annotation pipeline.}
To construct point-wise supervision at scale, we collect two types of candidate samples:
(i) candidate sets sampled from online recommendation models (each request contains multiple candidates); and
(ii) online exposure logs, including ads that users clicked or converted, as well as exposed but not clicked ads. 
These two sources complement each other: online candidate sets cover the model-induced exposure distribution, while exposure logs provide behavior-grounded positives and hard negatives.
Each instance is formatted into a structured prompt that contains the request context and the ad content, and a strong LLM is used to generate initial discrete aspect scores. Human annotators then verify and correct the scores to ensure label quality.
Verification feedback is further used to refine the prompt templates, forming a closed-loop pipeline:
\emph{automatic initial labeling $\rightarrow$ human verification $\rightarrow$ prompt iteration}.

\paragraph{Aspect scorer training.}
We first train PSJ by supervised fine-tuning (SFT) on human-verified point-wise samples to learn structured aspect prediction.
To improve consistency and robustness, we further apply GRPO with a KL constraint to an SFT reference model.
Let $y$ denote the predicted aspect-score sequence and $y^\ast$ the human label.
We define the dimension-wise reward as the sum of (i) an exact-match accuracy term $r_{\mathrm{acc}}^{d}$ and (ii) an order-consistency term $r_{\mathrm{ord}}^{d}$ that preserves the relative ordering induced by annotated scores.
The total point-wise reward aggregates over dimensions:
\begin{equation}
r_{\mathrm{asp}}(x,a,y)
=
\sum_{d\in D}\omega_d \Big(r_{\mathrm{acc}}^{d}(y,y^\ast) + r_{\mathrm{ord}}^{d}(y,y^\ast)\Big),
\quad |D|=4,\ \omega_d=1.
\end{equation}
GRPO is then employed to maximize the expected aspect reward $r_{asp}$ as~\cite{DeepSeekR1-Nature-2025}. The trained scorer is then used to generate aspect evidence $\mathbf{s}(x,a)$, which serves as input to the preference aggregation stage.

\subsubsection{Part B: User-conditional Preference Aggregation}
\vspace{4pt}

\paragraph{Motivation.}
While Part~A provides interpretable aspect evidence, industrial recommendation ultimately requires a \emph{single} holistic preference signal for ranking and selection.
Moreover, users may trade off semantic aspects differently under different request contexts.
Therefore, Part~B learns a user-conditional aggregation rule that integrates multi-dimensional aspect scores into a unified preference score.
We adopt \emph{pairwise} supervision for this stage because recommendation feedback and ranking objectives are inherently comparative: annotators can more consistently decide which candidate is better overall under the same context than assign calibrated absolute holistic scores, and such comparisons directly expose cross-aspect trade-offs.

\paragraph{Aggregation model.}
Given a user input $x$ and a candidate ad $a$, the aspect scorer from Part~A produces an aspect evidence vector $\mathbf{s}(x,a)\in\mathbb{R}^{4}$.
The aggregation model predicts user-dependent importance weights $\mathbf{w}(x)=g_\psi(x)\in\mathbb{R}^{4}$, where $w_d(x)\ge 0$ and $\sum_{d=1}^{4} w_d(x)=1$, and computes the holistic semantic score as
\begin{equation}
s_{\mathrm{hol}}(x,a)
=
\mathbf{w}(x)^\top \mathbf{s}(x,a)
=
\sum_{d=1}^{4} w_d(x)\, s_d(x,a).
\label{eq:hsj_aggregation}
\end{equation}

\paragraph{Pairwise supervision and annotation pipeline.}
We construct pairwise preference samples of the form $(x,a_i,a_j,y_{ij})$, where
$y_{ij}\in\{a_i\succ a_j,\ a_j\succ a_i\}$ indicates which ad is preferred under the same user input $x$.
Pairwise instances are collected from two complementary sources:
(i) \textit{intra-request comparisons} by sampling candidates within the same request, and
(ii) \textit{behavioral contrasts} from exposure logs that compare clicked/converted ads against exposed-but-not-clicked ads.
We follow an LLM-assisted human verification pipeline similar to Part~A: a strong LLM first provides an initial comparison judgment, and human annotators verify and correct labels to ensure quality.
Annotation rules are summarized in Table~\ref{tab:pairwise_annotation_rules}.

\paragraph{Preference optimization via group-based GRPO.}
For each labeled pair $(x,a_i,a_j)$ with $a_i \succ a_j$, we encourage $s_{\mathrm{hol}}(x,a_i) > s_{\mathrm{hol}}(x,a_j)$ and optimize the aggregation policy with GRPO-style updates. 
Concretely, for each $x$ we sample a \emph{group} of $G$ importance-level vectors
$\{{\mathbf{w}}^{(g)}(x)\}_{g=1}^{G}$ from the policy $\pi_\psi(\cdot\mid x)$, where
${\mathbf{w}}^{(g)}(x)\in\{0,\ldots,K\}^{4}$.
We compute holistic semantic scores by Eq.~\eqref{eq:hsj_aggregation} and define a preference-consistency reward:
\begin{equation}
r_{\mathrm{pw}}^{(g)}(x,a_i,a_j)
=
\mathbb{I}\big[s_{\mathrm{hol}}^{(g)}(x,a_i) > s_{\mathrm{hol}}^{(g)}(x,a_j)\big]
\quad \text{for}\quad a_i \succ a_j,
\label{eq:hsj_pair_reward}
\end{equation}
where $s_{\mathrm{hol}}^{(g)}(x,a)\triangleq (\mathbf{w}^{(g)}(x))^\top \mathbf{s}(x,a)$.
By sampling multiple weighting actions under the same $x$, we compute advantages $A_{\mathrm{pw}}^{(g)}$ in the corresponding rollout group to stabilize learning under the binary preference reward. Finally, we optimize a KL-regularized GRPO objective to improve preference consistency while keeping the policy close to a reference model:
\begin{equation}
\begin{aligned}
\max_\psi\ 
\mathbb{E}_{(x,a_i,a_j)\sim\mathcal{D}_{\mathrm{pw}}}
\Bigg[
&\ \mathbb{E}_{\{\hat{\mathbf{z}}^{(g)}\}_{g=1}^{G}\sim \pi_\psi(\cdot\mid x)}
\left[\frac{1}{G}\sum_{g=1}^{G} A_{\mathrm{pw}}^{(g)}\right]\\
&\ -\beta \,\mathrm{KL}\!\left(\pi_\psi(\cdot\mid x)\,\|\,\pi_{\mathrm{ref}}(\cdot\mid x)\right)
\Bigg].
\end{aligned}
\label{eq:hsj_grpo_weight}
\end{equation}

This training enables the aggregator to learn user-conditional cross-aspect trade-offs such that the resulting holistic semantic score $s_{\mathrm{hol}}(x,a)$ is consistent with human pairwise preferences.

\paragraph{Outputs for downstream optimization.}
PSJ finally provides (i) interpretable aspect scores $\mathbf{s}(x,a)$, and (ii) a holistic semantic score $s_{\mathrm{hol}}(x,a)$ obtained by weighting $\mathbf{s}(x,a)$ with predicted $\mathbf{w}(x)$.
The scalar reward $s_{\mathrm{hol}}(x,a)$ is used for subsequent recommendation optimization and will be fused with business rewards via A2PO.

\begin{figure}[t]
    \centering
    \includegraphics[width=\linewidth]{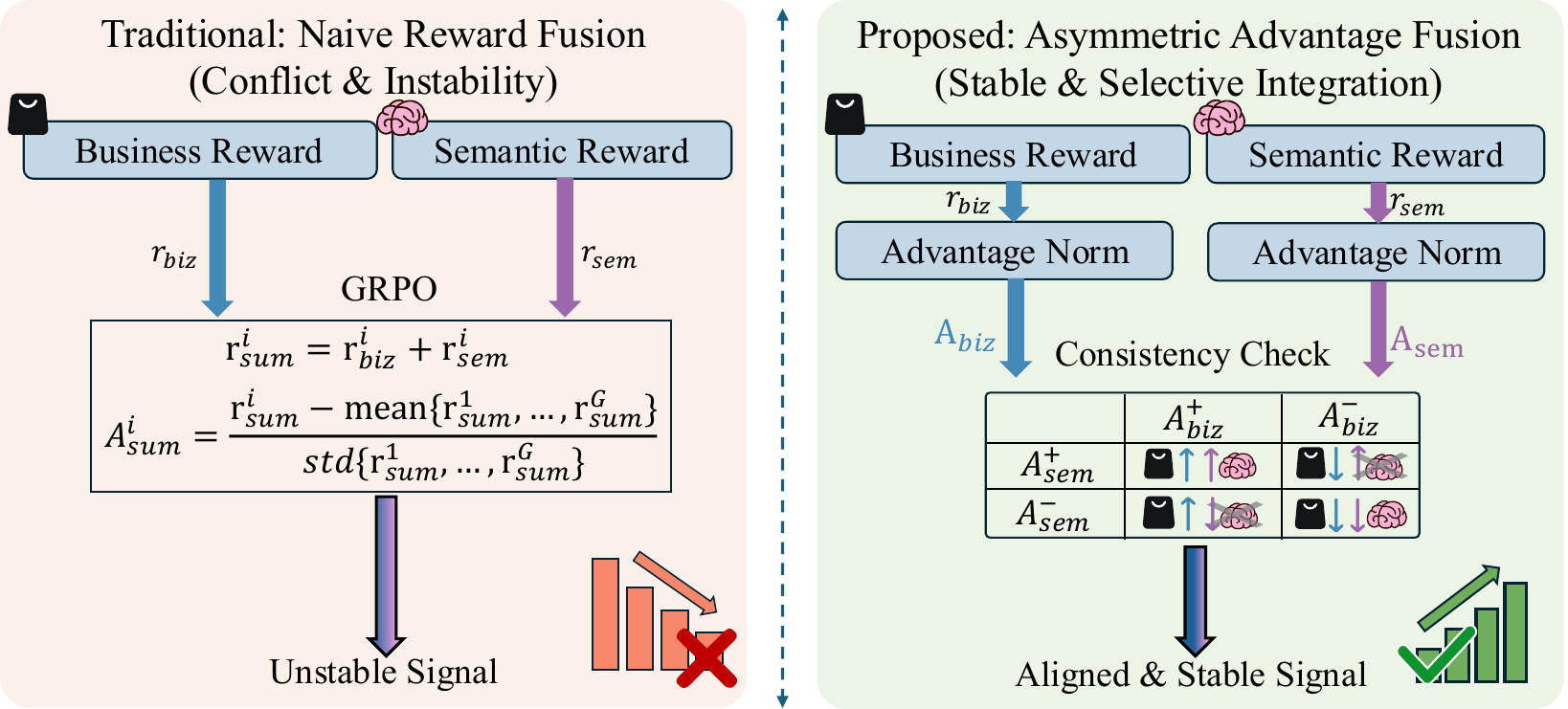}
    \caption{Asymmetric Advantage Fusion}
    \label{fig:a2po}
    \vspace{-0.2em}
\end{figure}

\subsection{Asymmetric Advantage Policy Optimization}
\label{subsec:a2po}


Industrial generative recommendation is often optimized against (i) business-driven objectives (e.g., eCPM, CTR, GMV) and (ii) fine-grained semantic preferences.
A naive approach is to fuse them at the reward level, e.g., $r = r_{\mathrm{biz}} + \alpha r_{\mathrm{sem}}$.
However, these rewards are heterogeneous in scale and distribution: business rewards are calibrated to platform metrics, while semantic rewards can exhibit context and judge-dependent variance.
As a result, reward-level fusion is brittle \cite{liu2026gdpo}, requiring sensitive manual tuning of $\alpha$ and often yielding unstable updates when one signal interferes with the other.
We therefore propose \textbf{Asymmetric Advantage Policy Optimization (A2PO)}, which performs fusion in the \emph{advantage space} and treats business optimization as an anchor while selectively injecting semantic guidance (see Figure~\ref{fig:a2po}).

\paragraph{Heterogeneous reward formulation.}
For each sampled candidate $y^{(i)}$ under input $x^{(i)}$, A2PO considers two rewards:
\begin{itemize}[leftmargin=*, itemsep=2pt]
    \item \textbf{Business reward} $r_{\mathrm{biz}}^{(i)}$: derived from platform objectives and feedback signals (e.g., eCPM/CTR/GMV-related rewards).
    \item \textbf{Semantic reward} $r_{\mathrm{sem}}^{(i)}$: computed by PSJ as a holistic semantic score $s_{\mathrm{hol}}(x^{(i)},a^{(i)})$ for mapped candidates $a^{(i)}=\mathrm{Map}(y^{(i)})$, capturing intent-relevance signals that are weak or missing in logs.
\end{itemize}
Rather than forcing these rewards into a shared numerical space, we standardize them within each rollout group of size $G$ and compute objective-specific advantages $A_{\mathrm{biz}}^{(i)}$ and $A_{\mathrm{sem}}^{(i)}$.

\paragraph{Asymmetric alignment in advantage space.}
Even after standardization, semantic advantages may be misaligned with business optimization due to judge uncertainty and open-domain ambiguity. To safeguard the platform's core revenue, A2PO uses business as an \textbf{anchor} and activates semantic guidance only when it is consistent with the business signal.
We define the fused advantage:
\begin{equation}
A^{(i)} = A_{\mathrm{biz}}^{(i)} + \lambda^{(i)} A_{\mathrm{sem}}^{(i)},
\label{eq:a2po_fused_adv}
\end{equation}
where the semantic coefficient $\lambda^{(i)}$ is computed by a dual-consistency rule:
\begin{equation}
\lambda^{(i)} =
\underbrace{\mathbb{I}\!\left(\mathrm{sign}\!\left(A_{\mathrm{biz}}^{(i)}\right)=\mathrm{sign}\!\left(A_{\mathrm{sem}}^{(i)}\right)\right)}_{\text{gate}}
\cdot
\underbrace{\frac{\min\!\left(\left|A_{\mathrm{biz}}^{(i)}\right|,\left|A_{\mathrm{sem}}^{(i)}\right|\right)}{\max\!\left(\left|A_{\mathrm{biz}}^{(i)}\right|,\left|A_{\mathrm{sem}}^{(i)}\right|\right)+\epsilon}}_{\text{magnitude}}.
\label{eq:a2po_lambda}
\end{equation}
Directional consistency prevents semantic guidance from pushing the policy against the business update direction, while magnitude balancing attenuates semantic influence when it is disproportionately stronger than the business signal.

\paragraph{Bounded semantic contribution.}
When $\lambda^{(i)}>0$, Eq.~\eqref{eq:a2po_lambda} implies
\begin{equation}
\left|\lambda^{(i)} A_{\mathrm{sem}}^{(i)}\right|
\le
\left|A_{\mathrm{biz}}^{(i)}\right|,
\label{eq:a2po_bound}
\end{equation}
thereby upper-bounding the effective semantic contribution by the business advantage magnitude.
This bound prevents semantic-induced spikes and stabilizes fusion without introducing additional tunable coefficients.

\paragraph{Optimization objective.}
We plug the rectified advantage $A^{(i)}$ into a GRPO-style clipped policy objective:
\begin{equation}
\begin{aligned}
\mathcal{L}_{\mathrm{A2PO}}
&=
\mathbb{E}_{i}\!\left[
\min\!\left(
\rho^{(i)} A^{(i)},
\ \mathrm{clip}\!\left(\rho^{(i)},1-\delta,1+\delta\right) A^{(i)}
\right)
\right],\\
\rho^{(i)}
&=
\frac{\pi_\theta\!\left(y^{(i)}\mid x^{(i)}\right)}{\pi_{\theta_{\mathrm{old}}}\!\left(y^{(i)}\mid x^{(i)}\right)}.
\end{aligned}
\label{eq:a2po_obj}
\end{equation}
where $\delta$ is the clipping threshold.
By rectifying semantic supervision through business-anchored consistency and a bounded semantic contribution, A2PO achieves stable multi-objective optimization under heterogeneous rewards while still benefiting from fine-grained semantic signals.

\section{Experiment}
We design experiments to answer the following questions:
\begin{itemize}[leftmargin=*, itemsep=2pt]
    \item \textbf{RQ1:} Does S-GRec improve recommendation performance over state-of-the-art generative methods?
    \item \textbf{RQ2:} Is PSJ's two-stage design necessary, and how does RL alignment affect aspect scoring quality?
    \item \textbf{RQ3:} Can A2PO resolve semantic--business objective conflicts via advantage-level fusion?
    \item \textbf{RQ4:} Does S-GRec remain effective under sparse semantic sampling, enabling cost-scalable training?
    \item \textbf{RQ5:} Does semantic supervision promote beyond-history exploration, and does it improve user experience?
    \item \textbf{RQ6:} Does S-GRec yield measurable business gains in online deployment?
\end{itemize}

\begin{table*}[ht]
\centering
\caption{Overall performance comparison on Amazon Industrial and Office datasets. Best results are \textbf{bolded} and second-best are \underline{underlined}.}
\label{tab:main_result}
\renewcommand{\arraystretch}{1.1}
\resizebox{\textwidth}{!}{%
\begin{tabular}{lcccccccccccc}
\toprule
\multirow{2}{*}{\textbf{Method}} &
\multicolumn{6}{c}{\textbf{Industrial}} &
\multicolumn{6}{c}{\textbf{Office}} \\
\cmidrule(lr){2-7}\cmidrule(lr){8-13}
& \textbf{HR@3} & \textbf{NDCG@3} & \textbf{HR@5} & \textbf{NDCG@5} & \textbf{HR@10} & \textbf{NDCG@10}
& \textbf{HR@3} & \textbf{NDCG@3} & \textbf{HR@5} & \textbf{NDCG@5} & \textbf{HR@10} & \textbf{NDCG@10} \\
\midrule
GRU4Rec & 0.0638 & 0.0542 & 0.0774 & 0.0598 & 0.0999 & 0.0669 & 0.0629 & 0.0528 & 0.0789 & 0.0595 & 0.1019 & 0.0669 \\
Caser & 0.0618 & 0.0514 & 0.0717 & 0.0555 & 0.0942 & 0.0628 & 0.0748 & 0.0615 & 0.0865 & 0.0664 & 0.1093 & 0.0737 \\
SASRec & 0.0790 & 0.0700 & 0.0909 & 0.0748 & 0.1088 & 0.0806 & 0.0861 & 0.0769 & 0.0949 & 0.0805 & 0.1120 & 0.0858 \\
HSTU & 0.0927 & 0.0885 & 0.1037 & 0.0918 & 0.1163 & 0.0958 & 0.1134 & 0.1031 & 0.1252 & 0.1079 & 0.1400 & 0.1126 \\
TIGER & 0.0852 & 0.0742 & 0.1010 & 0.0807 & 0.1321 & 0.0908 & 0.0986 & 0.0852 & 0.1163 & 0.0960 & 0.1408 & 0.1002 \\
LCRec & 0.0915 & 0.0805 & 0.1057 & 0.0862 & 0.1332 & 0.0952 & 0.0921 & 0.0807 & 0.1048 & 0.0859 & 0.1237 & 0.0920 \\
\midrule
BIGRec & 0.0931 & 0.0841 & 0.1092 & 0.0907 & 0.1370 & 0.0997 & 0.1069 & 0.0961 & 0.1204 & 0.1017 & 0.1434 & 0.1091 \\
D$^3$ & 0.1024 & 0.0991 & 0.1213 & 0.0989 & 0.1500 & 0.1082 & 0.1204 & 0.1055 & 0.1406 & 0.1139 & 0.1634 & 0.1213 \\
S-DPO & 0.1032 & 0.0906 & 0.1238 & 0.0991 & 0.1524 & 0.1082 & 0.1169 & 0.1033 & 0.1356 & 0.1110 & 0.1587 & 0.1255 \\
MiniOneRec & \underline{0.1143} & \underline{0.1011} & \underline{0.1321} & \underline{0.1084} & \underline{0.1586} & \underline{0.1167} & \underline{0.1217} & \underline{0.1088} & \underline{0.1420} & \underline{0.1172} & \underline{0.1634} & \underline{0.1242} \\
\ourmodel{} & \textbf{0.1160} & \textbf{0.1031} & \textbf{0.1359} & \textbf{0.1113} & \textbf{0.1632} & \textbf{0.1202} & \textbf{0.1319} & \textbf{0.1173} & \textbf{0.1496} & \textbf{0.1246} & \textbf{0.1689} & \textbf{0.1308} \\
\bottomrule
\end{tabular}
}
\end{table*}

\subsection{Experimental Setup}
\label{sec:setup}

\subsubsection{Datasets}
To ensure fair comparison, we follow the experimental setup of MiniOneRec~\cite{kong2025minionerec}. We evaluate on two widely used benchmarks from the Amazon Review dataset \cite{hou2024bridginglanguageitemsretrieval}: \textit{Office Products} (Office) and \textit{Industrial and Scientific} (Industrial). The datasets are pre-processed using the same scripts as MiniOneRec, filtering users and items with fewer than five interactions. Data splits and statistics are identical to those reported in \cite{kong2025minionerec}. Neither dataset contains temporal or location metadata, so the PSJ \textit{Context} aspect is dropped in all offline experiments.

\subsubsection{Baselines}
We compare S-GRec with two groups of baselines:
\begin{itemize}[leftmargin=*]
    \item \textbf{Non-LLM-based:} GRU4Rec \cite{GRU4Rec}, Caser \cite{Caser}, and SASRec \cite{SASRec} are traditional sequential models that rely on ID embeddings and sequence modeling. HSTU \cite{HSTU-ICML-2024}, TIGER \cite{TIGER-NeurIPS-2023}, and LC-Rec \cite{LC-Rec-ICDE-2024} adopt generative paradigms with Semantic IDs.
    \item \textbf{LLM-based:} BIGRec \cite{BIGRec-TORS-2025}, D$^3$ \cite{D3}, S-DPO \cite{kong2025sdpo}, and MiniOneRec \cite{kong2025minionerec}. MiniOneRec serves as our base model, on top of which S-GRec introduces PSJ and A2PO.
\end{itemize}

\subsubsection{Evaluation Metrics}
We report Hit Rate (HR@$K$) and Normalized Discounted Cumulative Gain (NDCG@$K$) for $K \in \{3, 5, 10\}$. All metrics are computed on the held-out test set.

\subsubsection{Implementation Details}
PSJ is built on \texttt{Qwen3-4B} \cite{yang2025qwen3} and trained following the two-stage pipeline in Section~\ref{sec:hsj}. We construct the training data from the production advertising system via a dedicated annotation platform: approximately 20k human-verified point-wise samples for aspect-level scoring and 40k pairwise preference samples for preference aggregation. On the Amazon benchmarks, PSJ scores candidates using item text metadata available in the dataset.
The generative recommender backbone is \texttt{Qwen2.5-1.5B} \cite{qwen2.5} with RQ-VAE tokenization, initialized from the MiniOneRec SFT checkpoint. A2PO training uses learning rate $1{\times}10^{-5}$, group size $G{=}16$, KL coefficient $\beta{=}0.04$. The instantiation of $r_{\mathrm{biz}}$ and $r_{\mathrm{sem}}$ follows the formulation in Section~\ref{subsec:a2po}. On the Amazon benchmarks, $r_{\mathrm{biz}}$ is instantiated as a ranking-based proxy derived from the ground-truth next item, since monetary signals are unavailable in public datasets; in the online deployment, $r_{\mathrm{biz}}$ corresponds to eCPM.

\subsection{Overall Performance (RQ1)}

Table~\ref{tab:main_result} presents the overall performance comparison. \ourmodel{} achieves the best results across all metrics on both datasets. Compared with its base model MiniOneRec, \ourmodel{} improves on all metrics, with representative gains of \textbf{2.9\%} HR@10 and \textbf{3.0\%} NDCG@10 on Industrial, and \textbf{3.4\%} HR@10 and \textbf{5.3\%} NDCG@10 on Office. 
The consistent improvements indicate that the semantic supervision from PSJ, fused through A2PO, provides complementary preference signals beyond what the base generative model captures from behavioral sequences alone, without modifying the serving-time architecture.

\subsection{Effectiveness of PSJ (RQ2)}

We validate PSJ from two angles: (i)~intrinsic scoring quality of the aspect-level scorer (Part~A) against human annotations, and (ii)~extrinsic impact of the two-stage pipeline (Part~A $\rightarrow$ Part~B) on downstream recommendation.
\begin{table}[t]
\centering
\caption{Intrinsic evaluation of Part~A aspect scoring quality.}
\label{tab:ablation_hsj}
\renewcommand{\arraystretch}{1.1}
\setlength{\tabcolsep}{3.5pt}
\small
\begin{tabular}{l|l|cc}
\toprule
\textbf{Variant} & \textbf{Description} & \textbf{PairAUC} & \textbf{PointAcc} \\
\midrule
DeepSeek-R1 & Zero-shot reference & 0.4617 & 0.4726 \\
\midrule
Qwen3-4B & Pre-trained only & 0.3987 & 0.4043 \\
\quad + SFT & Supervised fine-tuning & 0.4703 & 0.5202 \\
\quad \textbf{+ SFT + GRPO} & \textbf{+ RL alignment} & \textbf{0.8116} & \textbf{0.8687} \\
\bottomrule
\end{tabular}
\end{table}

\subsubsection{Aspect Scoring Quality}
We evaluate Part~A on 1{,}000 held-out point-wise samples from the production annotation platform, each containing a context--candidate pair with human scores across four semantic dimensions ($D{=}4$). We report two metrics, each computed per dimension and macro-averaged: \textbf{PairAUC}, the fraction of sample pairs whose predicted ordering agrees with the human-annotated ordering, and \textbf{PointAcc}, the rate at which the predicted discrete score exactly matches the human label.

Table~\ref{tab:ablation_hsj} includes DeepSeek-R1~\cite{DeepSeekR1-Nature-2025} as a zero-shot reference. The pre-trained Qwen3-4B underperforms R1 on both metrics. SFT on human-verified labels raises PairAUC by 18\% and PointAcc by 29\% relative to Qwen3-4B, showing that structured aspect prediction benefits substantially from domain supervision. GRPO alignment yields the most significant gain, boosting PairAUC and PointAcc to \textbf{0.8116} and \textbf{0.8687} respectively---roughly a 2$\times$ improvement over the pre-trained baseline. This confirms that domain-specific RL alignment is far more effective than model scale for aspect scoring.

\subsubsection{Two-Stage Decomposition Performance}
To test whether the two-stage PSJ design (aspect scoring $\rightarrow$ user-conditional aggregation) yields better semantic rewards than simpler alternatives, we fix A2PO and vary only the semantic reward source. Table~\ref{tab:ablation_hsj_rec} compares four configurations:
\begin{itemize}[leftmargin=*, itemsep=1pt]
    \item \textbf{MiniOneRec}: business reward only (no semantic signal).
    \item \textbf{+ Aspect only}: uniform aggregation of four aspect scores ($w_d{=}0.25$), bypassing Part~B.
    \item \textbf{+ Holistic only}: a single Qwen3-4B trained to output a scalar semantic score directly, without aspect decomposition.
    \item \textbf{\ourmodel{}}: PSJ with aspect scoring and user-conditional aggregation.
\end{itemize}

As shown in Table~\ref{tab:ablation_hsj_rec}, all semantic variants outperform the business-only baseline, indicating that PSJ provides complementary supervision. \textbf{+ Aspect only} improves HR@10 on Office/Industrial (1.9\%/1.7\% relative) but yields limited NDCG gains, suggesting that uniform weights fail to capture user-specific trade-offs. \textbf{+ Holistic only} further improves both metrics, yet \ourmodel{} achieves the best NDCG@10 on both datasets. On Industrial, \ourmodel{} slightly underperforms \textbf{Holistic only} on HR@10 while substantially improving NDCG@10, indicating that personalized aggregation primarily sharpens top-rank precision rather than increasing hit coverage. Overall, the decompose-then-aggregate design produces more effective semantic rewards than either stage alone.

\begin{table}[t]
\centering
\caption{Ablation of PSJ pipeline on downstream recommendation. All variants use A2PO for reward fusion.}
\label{tab:ablation_hsj_rec}
\renewcommand{\arraystretch}{1.05}
\setlength{\tabcolsep}{3pt}
\small
\begin{tabular}{l|cc|cc}
\toprule
\multirow{2}{*}{\textbf{Variant}} & \multicolumn{2}{c|}{\textbf{Office}} & \multicolumn{2}{c}{\textbf{Industrial}} \\
\cmidrule(lr){2-3}\cmidrule(lr){4-5}
& HR@10 & NDCG@10 & HR@10 & NDCG@10 \\
\midrule
MiniOneRec & 0.1634 & 0.1242 & 0.1586 & 0.1167 \\
+ Aspect only & 0.1665 & 0.1257 & 0.1613 & 0.1170 \\
+ Holistic only & 0.1677 & 0.1281 & 0.1637 & 0.1177 \\
\textbf{\ourmodel{}} & \textbf{0.1689} & \textbf{0.1308} & \textbf{0.1632} & \textbf{0.1202} \\
\bottomrule
\end{tabular}
\vspace{-1em}
\end{table}

\subsection{Effectiveness of A2PO (RQ3)}

We validate that A2PO's consistency-aware advantage-level fusion resolves the heterogeneous reward integration problem. Table~\ref{tab:ablation_a2po} organizes six variants into two groups: the first three rows compare fusion granularity (reward-level vs.\ advantage-level), while the last three ablate A2PO's internal submodules. All semantic-reward variants use PSJ.

\begin{table}[t]
\centering
\caption{Ablation of A2PO reward fusion. $r_{\mathrm{sem}}$: semantic reward enabled. Adv: advantage-level standardization.}
\label{tab:ablation_a2po}
\renewcommand{\arraystretch}{1.02}
\setlength{\tabcolsep}{0.95pt}
\small
\begin{tabular}{l|cccc|cc|cc}
\toprule
\multirow{2}{*}{\textbf{Variant}} & \multicolumn{4}{c|}{\textbf{Components}} & \multicolumn{2}{c|}{\textbf{Office}} & \multicolumn{2}{c}{\textbf{Industrial}} \\
\cmidrule(lr){2-5}\cmidrule(lr){6-7}\cmidrule(lr){8-9}
& $r_{\mathrm{sem}}$ & Adv & Gate & Wt. & HR@10 & NDCG@10 & HR@10 & NDCG@10 \\
\midrule
MiniOneRec & -- & -- & -- & -- & 0.1634 & 0.1242 & 0.1586 & 0.1167 \\
Reward-Sum & \checkmark & -- & -- & -- & 0.0904 & 0.0761 & 0.0585 & 0.0543 \\
Adv-Sum & \checkmark & \checkmark & -- & -- & 0.1515 & 0.1223 & 0.1348 & 0.1020 \\
\midrule
w/o Gate & \checkmark & \checkmark & -- & \checkmark & 0.1691 & 0.1302 & 0.1591 & 0.1169 \\
w/o Magnitude & \checkmark & \checkmark & \checkmark & -- & 0.1652 & 0.1294 & 0.1624 & 0.1189 \\
\textbf{\ourmodel{}} & \checkmark & \checkmark & \checkmark & \checkmark & \textbf{0.1689} & \textbf{0.1308} & \textbf{0.1632} & \textbf{0.1202} \\
\bottomrule
\end{tabular}
\end{table}

\subsubsection{Reward-level vs. Advantage-level Fusion}
Reward-Sum causes a catastrophic collapse---HR@10 drops by 44.6\% on Office and 63.1\% on Industrial relative to MiniOneRec---demonstrating that raw reward summation allows the semantic signal to dominate and destabilize the ranking objective. Adv-Sum, which standardizes each reward within its rollout group before summation, substantially recovers performance, validating that operating in the advantage space removes absolute-scale dependence. However, Adv-Sum still underperforms MiniOneRec on both datasets, indicating that unfiltered semantic advantages can still conflict with the business optimization direction. Empirically, the directional consistency rate between $A_{\mathrm{biz}}$ and $A_{\mathrm{sem}}$ is 45.4\% on the Amazon benchmarks and 49.8\% in the online advertising system, confirming that conflicts are frequent and that consistency gating is necessary.

\subsubsection{Ablation of A2PO Submodules}

Starting from Adv-Sum, adding either consistency gating or magnitude balancing already surpasses MiniOneRec on all metrics, indicating that each component can mitigate the residual conflict after standardization. Combining both yields the best NDCG@10 on both datasets, with 5.3\% and 3.0\% relative improvement over MiniOneRec on Office and Industrial respectively. Among A2PO variants, removing the gate hurts Industrial NDCG@10 the most, confirming that gating is critical for filtering directionally misaligned semantic signals.

\vspace{-0.5em}
\subsection{Cost-Scalability Analysis (RQ4)}

\begin{figure}[t]
\centering
\includegraphics[width=0.5\textwidth]{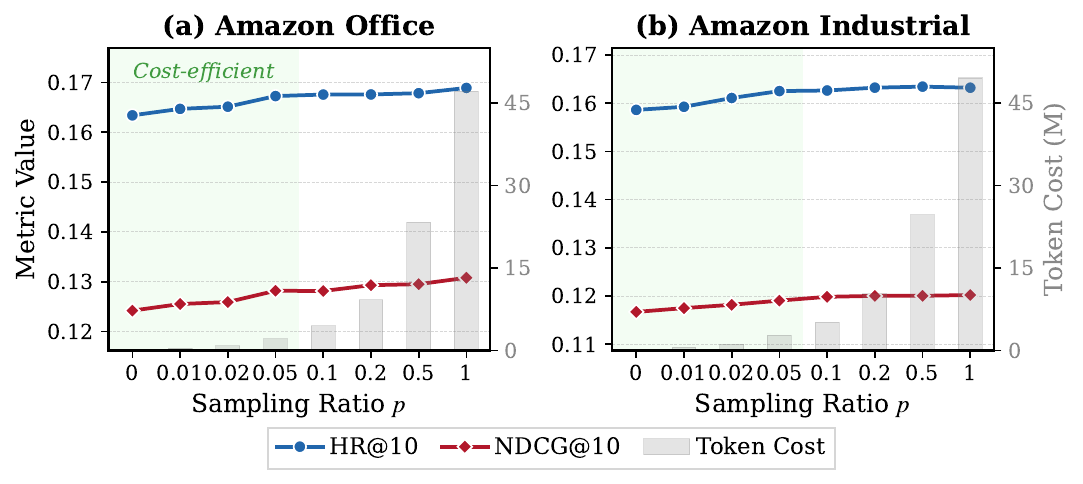}
\caption{HR@10 and NDCG@10 vs.\ semantic sampling ratio $p$ on Office and Industrial.}
\label{fig:sampling_ratio_cost_performance}
\end{figure}

A key practical concern for LLM-based semantic supervision is cost: invoking PSJ on every training sample is prohibitively expensive at industrial scale. We vary the sampling ratio $p$---the fraction of instances receiving semantic rewards (cost $\propto p$). Figure~\ref{fig:sampling_ratio_cost_performance} shows that S-GRec achieves near-saturated performance at $p{=}0.05$: 99.1\% of full performance on Office and 99.6\% on Industrial, enabling a \textbf{20$\times$} cost reduction. We adopt $p{=}0.05$ in online deployment. We suspect 5\% is sufficient because semantic supervision acts as a shared calibration signal on top of an already strong base policy, and A2PO further increases its precision by gating out directionally misaligned semantic advantages, leading to fast saturation.

\begin{figure}[t]
    \centering
    \includegraphics[width=0.8\linewidth]{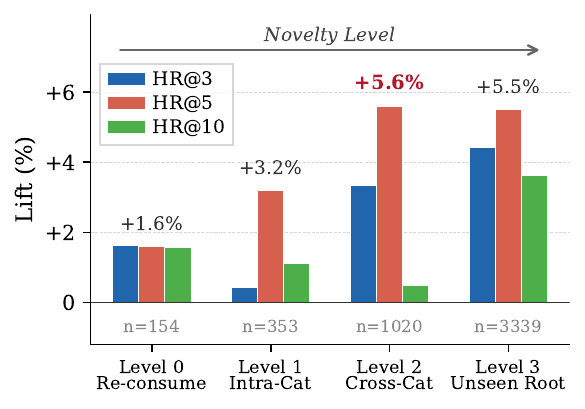}
    \caption{Relative HR lift (\%) of S-GRec over MiniOneRec across novelty levels on the Office test set. $n$ denotes the number of test samples per bucket.}
    \label{fig:history_novelty_lift}
\end{figure}

\begin{table*}[t]
\centering
\caption{Case study on Office. Left columns 1--3 are staggered to accommodate additional candidate analysis in right columns 4--11. $\bigstar$ denotes the ground-truth target.}
\label{tab:exploration_case_staggered}
\renewcommand{\arraystretch}{1.2}
\setlength{\tabcolsep}{6pt}
\footnotesize
\begin{tabular}{p{2.7cm} p{2.0cm} p{2.0cm} | l ccc c | cc}
\toprule
\textbf{User History} & \textbf{MiniOneRec } & \textbf{S-GRec } & \textbf{Item Candidate} & \textbf{Prof.} & \textbf{Fut.} & \textbf{Nov.} & $\boldsymbol{s_{\mathrm{hol}}}$ & \textbf{Mini} & \textbf{S-GRec} \\
\midrule
top5 & top5 & top5 & \textbf{Drafting Erasers} $\bigstar$ & 1.0 & \textbf{1.0} & \textbf{1.0} & \textbf{0.92} & \#6 & \textbf{\#1} \\
(1) Sharpener Blades & 1. Waterbrush & 1. \textbf{Erasers} $\bigstar$ & Metallic Pencils & 1.0 & 0.5 & 1.0 & 0.75 & \#3 & \#3 \\
(2) Brass Sharpener & 2. Pencils (48) & 2. Pencils (48) & Waterbrush (Small Tip) & 0.5 & 0.5 & 1.0 & 0.56 & \#1 & \#4 \\
(3) Artists' Pencils & 3. Metallic Pencils & 3. Metallic Pencils & Portrait Set (24) & 1.0 & 0.5 & 0.0 & 0.54 & \#4 & \#5 \\
(4) Inktense Pencils & 4. Portrait Set & 4. Waterbrush & Colored Pencils (48) & 1.0 & 0.0 & 0.0 & 0.38 & \#2 & \#2 \\
(5) Verithin Pencils & 5. Sargent Pencils & 5. Portrait Set & Sargent Pencils (12) & 0.5 & 0.5 & 0.0 & 0.35 & \#5 & -- \\
\bottomrule
\end{tabular}
\end{table*}

\subsection{Semantic-Guided Exploration Analysis (RQ5)}

The overall gains reported in RQ1 aggregate across all test instances. A natural question is: \emph{where does semantic supervision help most?} We hypothesize that its value concentrates on instances where behavioral signals are sparse---i.e., where the target item falls outside the user's historical patterns. To test this, we stratify the Amazon Office test set by history novelty and compare S-GRec against the MiniOneRec baseline, reporting relative HR lift.

\subsubsection{History Novelty Analysis}
We measure novelty w.r.t.\ the user's history $\mathcal{H}$ and bucket each target item by how far it generalizes beyond previously seen categories:
\begin{itemize}[leftmargin=*]
    \item \textbf{Level 0: Re-consumption.} The target item has appeared in history ($\mathit{target} \in \mathcal{H}$).
    \item \textbf{Level 1: Intra-Category.} The target is new, but shares the same sub-category $(c_1,c_2)$ with at least one historical item.
    \item \textbf{Level 2: Cross-Category.} The target shares the same root category $c_1$ with history, but belongs to a new sub-category $c_2$.
    \item \textbf{Level 3: Unseen Root.} The target's root category is unseen in history ($c_1(\mathit{target}) \notin c_1(\mathcal{H})$).
\end{itemize}

As shown in Figure~\ref{fig:history_novelty_lift}, the relative lift of S-GRec over the MiniOneRec baseline generally increases from Level~0 to Level~3. On \textit{Re-consumption} targets, S-GRec yields marginal gains (HR@5: \textbf{+1.6\%}), while on \textit{Unseen Root} targets the gain reaches \textbf{+5.5\%}. This trend suggests that high-novelty instances are harder for the behavior-only baseline: when targets deviate from historical categories, behavioral co-occurrence becomes sparse and MiniOneRec over-exploits familiar categories, under-ranking cross-category complements. PSJ supplies a transferable content-level quality prior that compensates for this failure mode, so the benefit of semantic supervision increases with novelty.

\subsubsection{Case Study}
Table~\ref{tab:exploration_case_staggered} illustrates how S-GRec enables cross-category exploration (Case \#1718). For a user purchasing art pencils, the ground-truth is a complementary drafting eraser---a category absent from their history. While the baseline ranks the target at \#6 (\textbf{miss@5}), S-GRec promotes it to \#1 (\textbf{hit@5}). 

We apply PSJ post-hoc to all candidates to interpret the ranking shift. The learned aspect weights for this user are $w_{\mathrm{Prof}}{=}0.15$, $w_{\mathrm{Fut}}{=}0.42$, $w_{\mathrm{Nov}}{=}0.31$ (Context is dropped; see Section~\ref{sec:setup}), heavily favoring Future Interest. History-similar candidates (e.g., Colored Pencils) score high on \textit{Profile} but low-to-zero on \textit{Future Interest} and \textit{Novelty}, yielding moderate $s_{\mathrm{hol}}$. Conversely, the target eraser achieves the highest $s_{\mathrm{hol}}$ (0.92) thanks to max \textit{Future Interest} and full \textit{Novelty}. The baseline's top-ranked Waterbrush, though novel, scores only $s_{\mathrm{hol}}{=}0.56$ due to moderate \textit{Future Interest} (0.5). Accordingly, S-GRec demotes the Waterbrush from \#1 to \#4 while elevating the eraser to the top, demonstrating that A2PO effectively distills content-level reasoning (pencils $\rightarrow$ erasers) into ranking improvements that behavioral co-occurrence fails to capture.

\vspace{-1em}
\subsection{Online A/B Test (RQ6)}
To evaluate the real-world impact of S-GRec, we first validate it in a simulation environment, where it achieves a \textbf{5.5\%} uplift in eCPM. Following this validation, we launch an online A/B test on 5\% of the advertising traffic in WeChat Channels. Compared with the production baseline, S-GRec yields \textbf{+1.19\%} GMV, \textbf{+1.55\%} GMV-Normal (ads optimized for clicks or conversions, which accounts for the majority of total GMV), \textbf{+1.16\%} CTR, and \textbf{$-$2.02\%} dislike rate. The concurrent improvement in both business revenue and user satisfaction confirms that S-GRec delivers meaningful commercial gains in a high-concurrency industrial environment, and validates the LLM-as-Judge paradigm as a practical path to enhance generative recommendation without requiring LLM inference at serving time.

\section{Conclusion}
This paper studies how to leverage LLM semantic priors in industrial generative recommendation under two deployment constraints: strict alignment to platform business objectives and prohibitive LLM inference cost at scale. To this end, we propose \ourmodel, which distills LLM semantics into \emph{train-time} supervision via a Personalized Semantic Judge (PSJ) that aggregates aspect-level semantic evidence into user-conditional scores for each candidate, and optimizes the generator with Asymmetric Advantage Policy Optimization (A2PO) that anchors updates on business rewards while selectively injecting reliable semantic guidance in the advantage space. Experiments on public benchmarks and a large-scale advertising system, including online A/B tests, show that \ourmodel consistently improves recommendation quality and delivers measurable business gains without requiring serving-time LLM inference. Future work includes extending the judge to richer preference dimensions and improving robustness under distribution shifts and evolving platform objectives.

\bibliographystyle{ACM-Reference-Format}
\bibliography{main}


\appendix


\clearpage
\section{Summary of Mathematical Notations}

Table \ref{tab:notations} presents all the symbols referenced in the Methodology section.

\begin{table}[htbp]
    \centering
    \caption{Summary of Mathematical Notations.}
    \label{tab:notations}
    \small 
    \renewcommand{\arraystretch}{1.3}
    \begin{tabularx}{\columnwidth}{l|X}
        \toprule
        \textbf{Symbol} & \textbf{Description} \\
        \midrule
        \multicolumn{2}{c}{\textit{Problem Definition \& Preliminaries}} \\
        \midrule
        $x$ & Unified user input including profile, behavior history, and request context features. \\
        $y$ & Semantic ID (SID) sequence representing an item, $\mathbf{y}=(y_1,\ldots,y_T)$. \\
        $a$ & Candidate item or ad mapped from the SID sequence, $a=\mathrm{Map}(y)$. \\
        $\pi_{\theta}$ & Generative recommendation policy with parameters $\theta$. \\
        $\mathcal{Y}(x)$ & Set of sampled candidate sequences for input $x$ with group size $G$. \\
        \midrule
        \multicolumn{2}{c}{\textit{Personalized Semantic Judge (PSJ)}} \\
        \midrule
        $\mathbf{s}(x,a)$ & Fine-grained aspect score vector ($\in \mathbb{R}^4$) covering four semantic dimensions. \\
        $s_d(x,a)$ & Discrete score for the $d$-th semantic aspect, where $d \in \{1,\dots,4\}$. \\
        $\mathbf{w}(x)$ & Context-dependent importance weight vector for preference aggregation. \\
        $s_{\mathrm{hol}}(x,a)$ & Holistic semantic score computed as the weighted sum $\mathbf{w}(x)^\top \mathbf{s}(x,a)$. \\
        $r_{\mathrm{asp}}$ & Point-wise scalar reward for aspect scorer training (Part A). \\
        $r_{\mathrm{pw}}$ & Pairwise consistency scalar reward for preference aggregation (Part B). \\
        \midrule
        \multicolumn{2}{c}{\textit{Asymmetric Advantage Policy Optimization (A2PO)}} \\
        \midrule
        $r_{\mathrm{biz}}, r_{\mathrm{sem}}$ & Business-driven rewards (e.g., eCPM) and semantic rewards (derived from $s_{\mathrm{hol}}$). \\
        $A_{\mathrm{biz}}, A_{\mathrm{sem}}$ & Standardized group-relative advantages for business and semantic objectives. \\
        $A^{(i)}$ & Final fused asymmetric advantage for the $i$-th sample. \\
        $\lambda^{(i)}$ & Adaptive coefficient gating the semantic advantage contribution. \\
        $\mathcal{L}_{\mathrm{A2PO}}$ & The final objective function for S-GRec policy optimization. \\
        \bottomrule
    \end{tabularx}
\end{table}

\section{Data Annotation}
\label{app:data_annotation}
The point‑wise fine‑grained scoring rules are defined in Table \ref{tab:fine_grained_scoring_specs}, which includes four tasks: user‑profile relevance, future‑interest relevance, novelty, and contextual relevance. We employ DeepSeek‑R1 as the initial judge model to generate preliminary scores along with intermediate reasoning. Subsequently, we introduce a human calibration step that focuses on verifying whether the intermediate reasoning produced by DeepSeek‑R1 is correct and whether the final scores adhere to the defined scoring rules. At the same time, we collect feedback from human annotators to refine ambiguous aspects of the original annotation‑rule definitions.

The pair‑wise scoring rule is defined in Table \ref{tab:pairwise_annotation_rules}, which contains one task: given a user’s information and two candidate advertisements, determine which advertisement the user is more likely to prefer. We provide four options: "$A>B$" (i.e., the user prefers A over B), "$A<B$" (the user prefers B over A), "tie (both equally good)", and "tie (both equally poor)", so as to reduce labeling noise.

\begin{table*}[t]
\centering
\caption{Detailed Specifications for the Four Fine-grained Scoring Dimensions of PSJ}
\label{tab:fine_grained_scoring_specs}
\small
\begin{tabular}{p{0.4cm} p{2.5cm} p{6cm} p{6cm}}
\toprule
\textbf{ID} & \textbf{Criterion} & \textbf{Rules and Guidelines} & \textbf{Scoring Rules} \\
\midrule
0 & User Profile Relevance & Evaluate whether the recommendation aligns with the user's long-term, stable profile.
\begin{itemize}[leftmargin=*,topsep=0pt, itemsep=-1pt, parsep=2pt]
    \item[\textbf{a.}] Behavior importance: \textbf{Ad interaction > Article reading > Video play.}
    \item[\textbf{b.}] Rank all interest tags by importance (strongest first), keep at most 10.
    \item[\textbf{c.}] Behavior weight: \textbf{Conversion > Click.}
    \item[\textbf{d.}] Summarize user's stable interests from basic info and history using LLM.
\end{itemize}
&
\begin{itemize}[leftmargin=*,topsep=0pt, itemsep=-1pt, parsep=2pt]
    \item Score range: [-1, -0.5, 0, 0.5, 1] (5 levels).
    \item Aligns with strong interest: \textbf{1}; weak interest: \textbf{0.5}; no match: \textbf{0}.
    \item Contradicts profile (strong): \textbf{-1}; (mild): \textbf{-0.5}. (e.g., do not recommend degree-upgrading ads to highly-educated users).
\end{itemize} \\
\midrule
1 & Future Interest Relevance & Assess if the recommendation matches the user's likely next interest.
\begin{itemize}[leftmargin=*,topsep=0pt, itemsep=-1pt, parsep=2pt]
    \item[\textbf{a.}] Enrich/expand item attributes/content for better inference.
    \item[\textbf{b.}] Infer next possible clicks based on temporal continuity and interest trends. Ignore isolated noisy behaviors.
    \item[\textbf{c.}] Explicitly exclude ads violating user interests or common sense.
    \item[\textbf{d.}] Predict the user's most likely next interest point for clicking.
\end{itemize}
&
\begin{itemize}[leftmargin=*,topsep=0pt, itemsep=-1pt, parsep=2pt]
    \item Score range: [0, 0.5, 1] (3 levels).
    \item Aligns with strong predicted interest: \textbf{1}; weak: \textbf{0.5}; no match: \textbf{0}.
    \item Example: If history suggests buying hiking gear, relevant outdoor ads score higher.
\end{itemize} \\
\midrule
2 & Novelty & Judge if the result satisfies both conditions: 1) aligns with user interest; 2) similar ads not clicked in user history. &
\begin{itemize}[leftmargin=*,topsep=0pt, itemsep=-1pt, parsep=2pt]
    \item Score range: [0, 0.5, 1] (3 levels).
    \item \textbf{0}: No novelty; \textbf{0.5}: Weak novelty; \textbf{1}: Strong novelty.
\end{itemize} \\
\midrule
3 & Contextual Relevance & Evaluate match with user's situational context: location, time, season, age, etc. &
\begin{itemize}[leftmargin=*,topsep=0pt, itemsep=-1pt, parsep=2pt]
    \item Score range: [-1, 0] (2 levels).
    \item \textbf{-1}: Contradicts context (penalize); \textbf{0}: Matches or irrelevant.
    \item Exclude temporally irrelevant ads (e.g., post-618 sale ads).
    \item Match based on season, holidays, current time.
\end{itemize} \\
\bottomrule
\end{tabular}
\end{table*}

\begin{table*}[t]
\centering
\caption{Annotation Rules and Output Format for the Pair-wise Holistic Scoring Task}
\label{tab:pairwise_annotation_rules}
\small
\begin{tabular}{p{0.6cm} p{2.8cm} p{8.5cm} p{3.5cm}}
\toprule
\textbf{ID} & \textbf{Criterion} & \textbf{Rules and Constraints} & \textbf{Output Format} \\
\midrule
0 & \textbf{Core Task:} Given two candidate ads, determine which one the user is more likely to click. &
\begin{itemize}[leftmargin=*,topsep=0pt, itemsep=-1pt, parsep=2pt]
    \item[\textbf{a.}] Prioritize behavior types: \textbf{Ad interaction importance > Article reading importance > Video play importance.}
    \item[\textbf{b.}] Consider temporal continuity of user behavior; reason about the next potential interest.
    \item[\textbf{c.}] For each ad, analyze alignment with the user's next interest and assess the degree of alignment.
    \item[\textbf{d.}] Conduct a thorough comparison between the two candidate ads.
    \item[\textbf{e.}] If ads are very similar and hard to distinguish, indicate lower certainty and set ranks for both to 0.
    \item[\textbf{f.}] If both ads are of very low quality and clearly irrelevant to user interest, set ranks for both to 0.
\end{itemize}
&
\begin{itemize}[leftmargin=*,topsep=0pt, itemsep=-1pt, parsep=2pt]
    \item \textbf{Ranking result:} Winning ad is marked as \textbf{1}, the other as \textbf{0}.
    \item \textbf{Degree of preference:} Score indicating how much better the winning ad is (0: Identical, 5: Moderate, 10: Very Obvious).
\end{itemize} \\
\bottomrule
\end{tabular}
\end{table*}

\section{Prompts}
\label{app:prompts}
\begin{figure*}[htbp]
    \centering
    \includegraphics[width=1\textwidth]{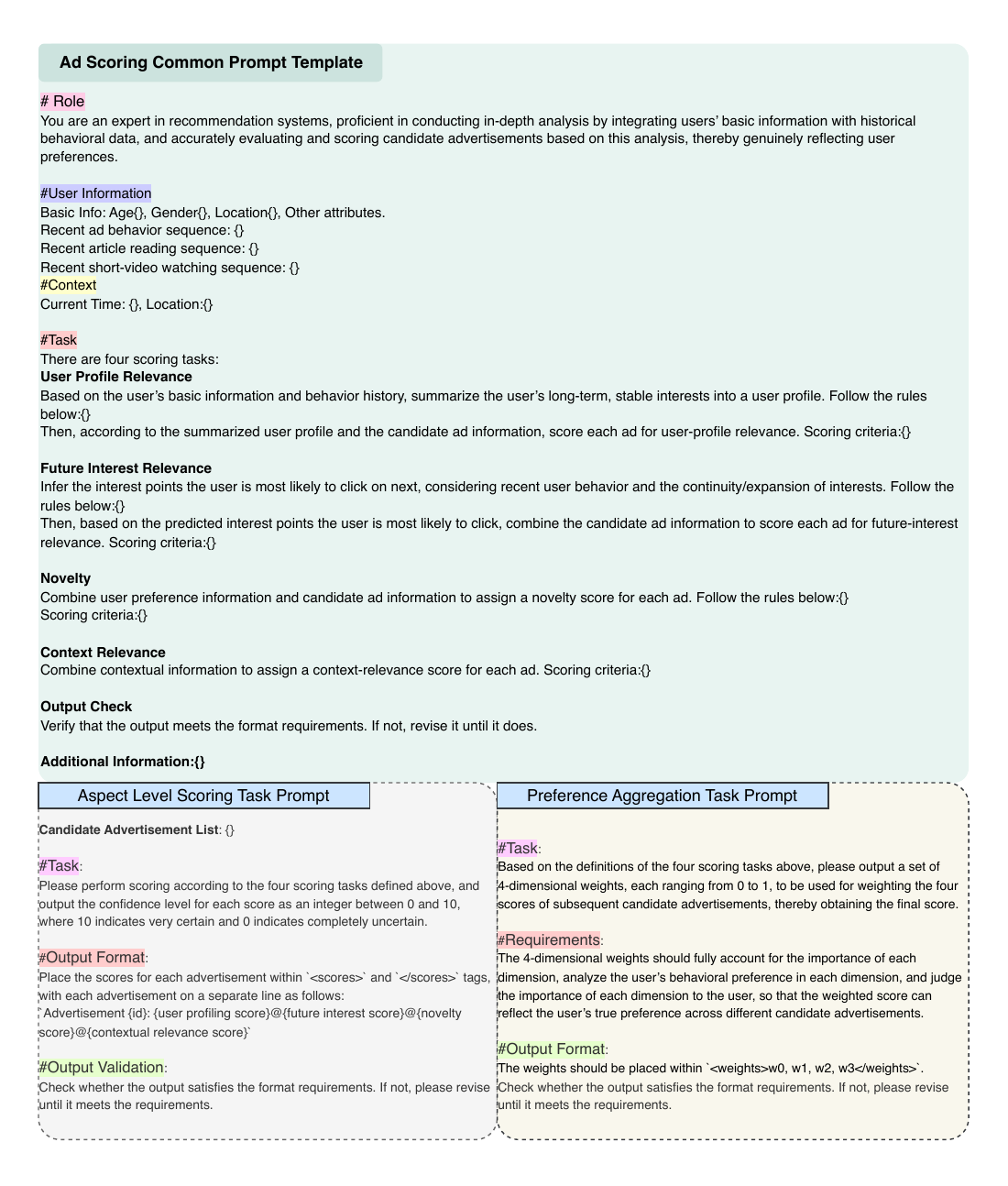}
    \caption{Prompt templates for PSJ. The common prompt provides user information and scoring task definitions, while the task-specific prompts define aspect-level scoring and preference aggregation respectively.}
    \label{fig:prompt_common}
\end{figure*}

As depicted in Figure \ref{fig:prompt_common}, the aspect-level semantic scoring task and the user-conditional preference aggregation task can share the same user input. Therefore, we structure the prompt template into two parts: a common prompt defined in \textit{Ad Scoring Common Prompt Template}, which contains the complete user information and the definitions of the four scoring tasks (with detailed scoring rules for each task provided in Table \ref{tab:fine_grained_scoring_specs}), and a task‑specific part that differs for each objective.

For aspect-level semantic scoring task, the template is defined in \textit{Aspect Level Scoring Task Prompt}, while for user-conditional preference aggregation task, the template is defined in \textit{Preference Aggregation Task Prompt}.

\section{Case Study}
\begin{figure*}[htbp]
    \centering
    \includegraphics[width=1\textwidth]{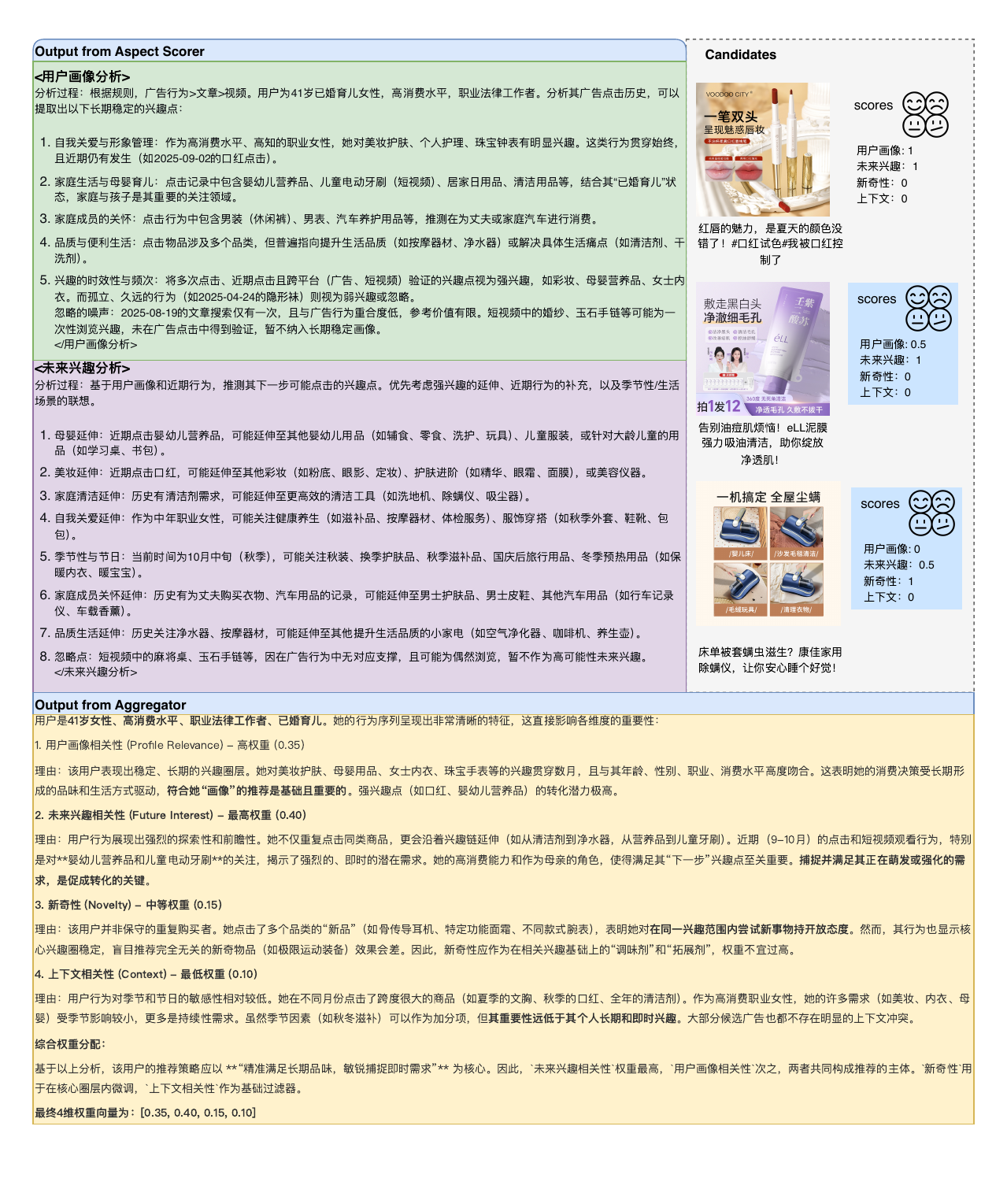}
    \caption{Case study of PSJ scoring process. The aspect-level scorer analyzes user behavioral history to produce fine-grained scores, and the aggregator infers user-conditional dimension weights.}
    \label{fig:case}
\end{figure*}

Figure \ref{fig:case} presents a case of the scoring process. For aspect-level scoring, the LLM conducts in‑depth analysis of the user’s behavioral history to extract user‑profile information and future‑interest signals. The LLM is capable of accurately identifying the user’s interest preferences and predicting potential interest extensions.

The Aggregator LLM performs reasoning on the user's behavioral patterns across the four defined dimensions, assesses the influence of each dimension on the user's behavior, and thereby designs a reasonable set of scoring weights.

\section{Deployment}
\begin{figure}[H]
    \centering
    \includegraphics[width=0.45\textwidth]{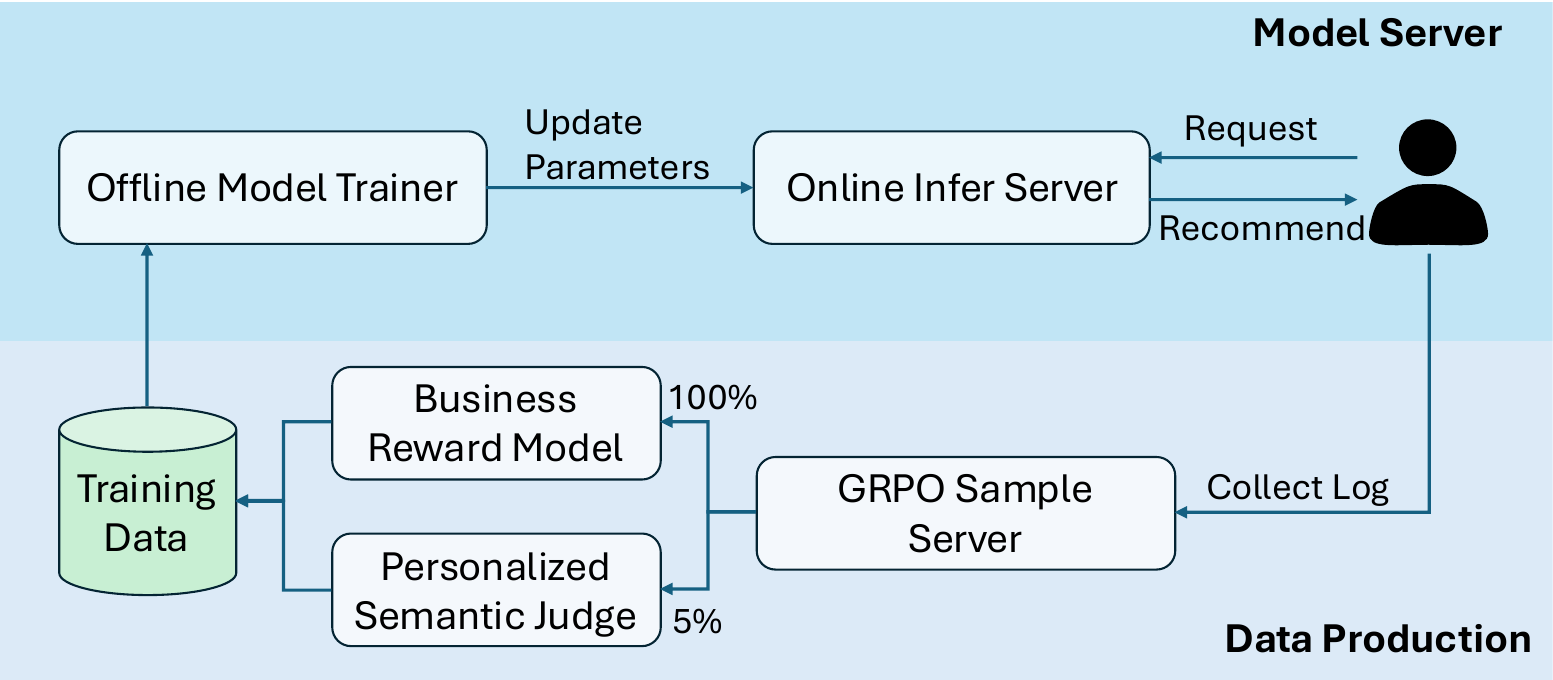}
    \caption{Framework of S-GRec Deployment}
    \label{fig:deployment}
\end{figure}

S-GRec has been successfully deployed in the WeChat Channels service at Tencent. As shown in Figure~\ref{fig:deployment}, the architecture comprises the Model Server and Data Production modules. During operation, the Online Infer Server processes user requests and transmits interaction logs to the GRPO Sample Server. We employ a resource-efficient sampling strategy for reward generation: the Business Reward Model scores 100\% of the data for general business metrics, while the Personalized Semantic Judge evaluates a 5\% subset for semantic quality. The Offline Model Trainer utilizes the aggregated training data and periodically updates the online server parameters.

\end{document}